\documentclass[aps,twocolumn,showpacs]{revtex4}
\usepackage{amsmath}
\usepackage{epsfig}

\begin{document}

\title{Investigating the $p$-$\Omega$ Interaction and Correlation Functions}
\author{Ye Yan$^1$}\email{221001005@njnu.edu.cn}
\author{Qi Huang$^1$}\email{06289@njnu.edu.cn}
\author{Youchang Yang$^2$}\email{yangyc@gues.edu.cn}
\author{Hongxia Huang$^1$}\email{hxhuang@njnu.edu.cn(Corresponding author)}
\author{Jialun Ping$^1$}\email{jlping@njnu.edu.cn}
\affiliation{$^1$Department of Physics, Nanjing Normal University, Nanjing 210023, China}
\affiliation{$^2$School of Science, Guizhou University of Engineering Science, Bijie 551700, China}

\begin{abstract}
Motivated by experimental measurements, we investigate the $p$-$\Omega$ correlation functions and interactions on the basis of a quark model.
By solving the inverse scattering problem with channel coupling, we renormalize the coupling to other channels into an effective single-channel $p$-$\Omega$ potentials.
The effects of Coulomb interaction and spin-averaging are also discussed.
According to our results, the depletion of the $p$-$\Omega$ correlation functions, which is attributed to the $J^P = 2^+$ bound state not observed in the ALICE Collaboration's measurements [Nature \textbf{588}, 232 (2020)], can be explained by the contribution of the attractive $J^P = 1^+$ component in spin-averaging.
So far, we have provided a consistent description of the $p$-$\Omega$ system from the perspective of the quark model, including the energy spectrum, scattering phase shifts, and correlation functions.
The existence of the $p$-$\Omega$ bound state has been supported by all three aspects.
Additionally, a sign of the $p$-$\Omega$ correlation function's subtle sub-unity part can be seen in experimental measurements, which warrants more precise verification in the future.

\end{abstract}


\maketitle

\setcounter{totalnumber}{5}

\section{Introduction}
\label{sec1}
Understanding hadron-hadron interactions is a cornerstone of modern nuclear and particle physics.
Insights gained from hadron-hadron interactions research contribute to our understanding of Quantum Chromodynamics (QCD) and help explore the properties of matter at the smallest scales.
To study these interactions, scattering hadrons off each other~\cite{Eisele:1971mk,Alexander:1968acu} is an important and effective approach.
By examining the scattering processes, people can gain valuable insights into the forces that govern hadronic interactions.
However, high-quality measurements such as scattering processes, are not suitable for unstable particles in experimental studies.
Therefore, femtoscopic correlations between hadron pairs in momentum space have become a powerful tool for experimentally studying the hadron-hadron interaction~\cite{STAR:2014dcy,ALICE:2020mfd,STAR:2015kha,ALICE:2018nnl,ALICE:2019hdt,ALICE:2019gcn,Fabbietti:2020bfg,ALICE:2021cpv,ALICE:2021njx,ALICE:2022enj}.
For instance, the ALICE Collaboration investigated the $p$-$\Omega$ and the $p$-$\Xi$ interactions through correlation functions~\cite{ALICE:2020mfd}.
The measurements of the proton-deuteron and deuteron-deuteron correlation functions have also been reported recently~\cite{STAR:2024lzt}.
The collider experiments have greatly advanced the field of particle physics, driving significant breakthroughs in our understanding~\cite{CMS:2024pts,LHCb:2024tpv,Belle:2012iwr,Shou:2024uga,Chen:2024zwk,ATLAS:2024wla}.
In addition, the study of correlation functions of various systems has also achieved substantial results in theoretical works~\cite{Ohnishi:1998at,Morita:2014kza,Ohnishi:2016elb,Hatsuda:2017uxk,Haidenbauer:2018jvl,Kamiya:2019uiw,Haidenbauer:2020kwo,Ohnishi:2021ger,Ogata:2021mbo,Mrowczynski:2021bzy,Graczykowski:2021vki,Kamiya:2021hdb,Haidenbauer:2021zvr,Liu:2022nec,Liu:2023uly,Liu:2023wfo,Molina:2023oeu,Liu:2024nac,Vidana:2023olz,Sarti:2023wlg,Li:2024tof,Molina:2023jov,Feijoo:2024bvn,Albaladejo:2023wmv,Feijoo:2023sfe,Ikeno:2023ojl,Kamiya:2024diw,Albaladejo:2023pzq,Torres-Rincon:2023qll,Abreu:2024qqo,Li:2024tvo,Albaladejo:2024lam,Etminan:2024uvc,Jinno:2024rxw,Etminan:2024nak}.

Moreover, of great significance in studying hadron-hadron interactions is to test whether two hadrons can form exotic hadronic states.
In recent years, significant progress has been made in the study of exotic states~\cite{Liu:2013waa,Guo:2017jvc,Liu:2019zoy,Hyodo:2020czb,Chen:2021ftn,Chen:2022asf,Meng:2022ozq,Huang:2023jec,Liu:2024uxn,Liu:2021pdu}, including tetraquarks~\cite{LHCb:2020bwg,CMS:2023owd}, pentaquarks~\cite{LHCbPc2015,LHCbPc2019,LHCb:2020jpq,LHCb:2022ogu}, and dibaryons~\cite{Urey:1932pvy,Bashkanov:2008ih,WASA-at-COSY:2011bjg,WASA-at-COSY:2012seb,WASA-at-COSY:2014dmv,WASA-at-COSY:2014qkg,BESIII:2023vvr}.
For instance, recently the BESIII Collaboration reported the observation of new $X(1880)$ in the line shape of the $3(\pi^+ \pi^-)$ invariant mass spectrum~\cite{BESIII:2023vvr}, which is considered as evidence for the existence of a $p\bar{p}$ bound state.
Hence, the study of hadron-hadron interactions and femtoscopic correlation functions is interconnected and complementary to each other.

The $p$-$\Omega$ with $J^P = 2^+$ (also expressed as $N$-$\Omega$ in some works without considering the Coulomb interaction) is also considered a possible dibaryon state in theoretical studies and has generated a lot of interest.
It was first predicted by J. T. Goldman \textit{et al.} using two different quark models~\cite{Goldman:1987ma}.
Subsequently, it was pointed out that the $p$-$\Omega$ with $J^P = 2^+$ is more likely than that with $J^P = 1^+$ by employing a quark-cluster model~\cite{Oka:1988yq}.
The bound state $p$-$\Omega$ with $J^P = 2^+$ was also derived utlizing the chromomagnetic model~\cite{Silvestre-Brac:1992xsl}, with the predicted binding energy of about 33 MeV.
In the framework of the chiral quark model~\cite{Dai:2007gc} and the quark delocalization color screening model (QDCSM)~\cite{Huang:2015yza}, the $p$-$\Omega$ with $J^P = 2^+$ was predicted to be a weakly bound state.
In Ref.~\cite{Chen:2021hxs}, the results of QCD sum rules indicated that there may exist a $p$-$\Omega$ dibaryon bound state with $J^P = 2^+$ and a binding energy of about 21 MeV, while the mass of the state with $J^P = 1^+$ is above the corresponding threshold.

The $p$-$\Omega$ with $J^P = 2^+$ was also studied using the (2+1)-flavor lattice QCD simulations by the HAL QCD collaboration~\cite{HALQCD:2014okw},
they found a $J^P = 2^+$ $p$-$\Omega$ bound state with a binding energy of 18.9 MeV under the condition that $m_\pi = 875$ MeV.
Four years later, the HAL QCD collaboration re-studied the $J^P = 2^+$ $p$-$\Omega$ system with nearly physical quark masses ($m_\pi = 146$ MeV),
the updated binding energy, scattering length, and effective range were obtained as 1.54 MeV, 5.30 fm, and 1.24 fm, respectively~\cite{HALQCD:2018qyu}.

With the help of the lattice QCD simulation data, several theoretical studies were carried out.
Scattering lengths obtained by the HAL QCD collaboration were used in Refs.~\cite{Sekihara:2018tsb,Sekihara:2023ihc} to further investigate the properties of the $p$-$\Omega$ dibaryon using the constituent quark model.
The width of the $p$-$\Omega$ dibaryon with $J^P = 2^+$ is calculated to be 1.5 MeV~\cite{Sekihara:2018tsb} and 4.6 MeV~\cite{Sekihara:2023ihc}, and the two results correspond to the model parameters set by two sets of scattering data obtained by the HAL QCD collaboration.
Additionally, using a phenomenological Lagrangian approach, the sum of the $p$-$\Omega$ decay rates was predicted to be 166--682 keV~\cite{Xiao:2020alj}.
In Ref.~\cite{Zhang:2020dma}, using the $p$-$\Omega$ interaction potential by the lattice QCD simulation to obtain wave functions, the production of the $p$-$\Omega$ dibaryon was estimated by using of a dynamical coalescence mechanism.
The productions of $p$-$\Omega$ were also investigated with the help of the effective Lagrangian approach~\cite{Liu:2022uap} and a covariant coalescence model~\cite{Pu:2024kfh}, which can be helpful for future experimental searches.

Based on two lattice simulations for the $p$-$\Omega$ with $J^P = 2^+$~\cite{HALQCD:2014okw,HALQCD:2018qyu} and assuming that the $p$-$\Omega$ with $J^P = 1^+$ wave function is completely absorbed into octet-octet states, the $p$-$\Omega$ correlation functions were studied in Refs.~\cite{Morita:2016auo,Morita:2019rph}.
Furthermore, in Ref.~\cite{Morita:2016auo}, the ratio of correlation functions between small and large collision systems, \( C_{SL}(Q) \), is proposed as a new measure to extract the strong $p$-$\Omega$ interaction with minimal contamination from the Coulomb attraction.

The first measurement of the $p$-$\Omega$ correlation function in heavy-ion collisions at $\sqrt{s_{NN}} = 200$ GeV was reported by the STAR Collaboration and the results indicated that the scattering length is positive for the $p$-$\Omega$ interaction and favored the $p$-$\Omega$ bound state hypothesis~\cite{STAR:2018uho}.
In Ref.~\cite{ALICE:2020mfd}, the ALICE Collaboration reported the measurement of the $p$-$\Omega$ correlation in $p$ + $p$ collisions at $\sqrt{s} = 13$ TeV at the LHC.
In comparison to the results based on the lattice data~\cite{HALQCD:2018qyu,Morita:2019rph}, the depletion of the correlation function, visible in the calculations around $k$ = 150 MeV/$c$ due to the presence of a $p$-$\Omega$ bound state, is not observed in the measured correlations.

The QDCSM is also an effective method for dealing with hadron-hadron interactions.
The model gives a good description of $N$-$N$ and $Y$-$N$ interactions and the properties of the deuteron~\cite{Ping:2000dx,Ping:1998si,Wu:1998wu,Pang:2001xx}.
It is also employed to calculate the hadron-hadron scattering phase shifts and the exotic hadronic states~\cite{Xue:2020vtq,Yan:2023tvl,Huang:2013nba}.
In our previous work~\cite{Huang:2015yza}, We investigated the $p$-$\Omega$ dibaryon with $J^P = 2^+$ in the QDCSM and find a bound state.
Motivated by experimental measurements of the $p$-$\Omega$ correlation functions, we aim to extend the QDCSM to theoretically compute the correlation functions.
This expansion allows the model to provide a unified description of hadron-hadron interactions, encompassing energy spectra, scattering processes, and correlation functions.

In this work, utilizing the scattering phase shifts with channel coupling calculated by the QDCSM, we renormalize the coupling to other channels into an effective single channel $p$-$\Omega$ potential.
In calculating the $p$-$\Omega$ correlation functions, we discuss the effects of the Coulomb interaction and spin-averaging on the $p$-$\Omega$ correlation functions.
Next, considering the error in the source function, we compare the correlation function obtained from our model with those from lattice data and the latest experimental measurement.
In terms of energy spectrum, scattering phase shift, and correlation function, we have completed the consistent description of the $p$-$\Omega$ system and the existence of a $p$-$\Omega$ bound state with $J^P = 2^+$ is supported.

This paper is organized as follows.
In the next section, we provide an introduction to calculating the $p$-$\Omega$ correlation function and the Gel'fand-Levitan-Marchenko (GLM) method.
In section~\ref{sec3}, results and discussions of $p$-$\Omega$ correlation function are given, followed by a summary.

\section{Theoretical formalism}\label{sec2}

\subsection{Two-particle correlation function} \label{sec2.1}

Experimentally, the correlation function $C(\boldsymbol{k})$ can be measured based on:
\begin{align}
	C(\boldsymbol{k}) & = \xi(\boldsymbol{k}) \frac{N_{\text{same}}(\boldsymbol{k})}{N_{\text{mixed}}(\boldsymbol{k})},
\end{align}
where $N_{\text{same}}(\boldsymbol{k})$ and $N_{\text{mixed}}(\boldsymbol{k})$ represent the $\boldsymbol{k}$ distributions of hadron-hadron pairs produced in the same
and in different collisions, respectively, and $\xi(\boldsymbol{k})$ denotes the corrections for experimental effects.
In theoretical studies, the correlation function can be calculated using the Koonin$-$Pratt (KP) formula~\cite{Koonin:1977fh,Pratt:1990zq,Bauer:1992ffu}:
\begin{align}
	C(\boldsymbol{k}) & = \frac{N_{12}\left(\boldsymbol{p}_{1}, \boldsymbol{p}_{2}\right)}{N_{1}\left(\boldsymbol{p}_{1}\right) N_{2}\left(\boldsymbol{p}_{2}\right)} \\
	& \simeq \frac{\int \mathrm{d}^{4} x_{1} \mathrm{~d}^{4} x_{2} S_{1}\left(x_{1}, \boldsymbol{p}_{1}\right) S_{2}\left(x_{2}, \boldsymbol{p}_{2}\right)|\Psi(\boldsymbol{r}, \boldsymbol{k})|^{2}}{\int \mathrm{d}^{4} x_{1} \mathrm{~d}^{4} x_{2} S_{1}\left(x_{1}, \boldsymbol{p}_{1}\right) S_{2}\left(x_{2}, \boldsymbol{p}_{2}\right)} \\ \label{ignore}
	& \simeq \int \mathrm{d} \boldsymbol{r} S_{12}(r)|\Psi(\boldsymbol{r}, \boldsymbol{k})|^{2},
\end{align}
where $S_{i}(x_{i}, \boldsymbol{p}_{i})~(i = 1, 2)$ is the single particle source function of the hadron $i$ with momentum $\boldsymbol{p}_{i}$, $\boldsymbol{k} = (m_2 \boldsymbol{p}_{1} - m_1 \boldsymbol{p}_{2})/(m_1 + m_2)$ is the relative momentum in the center-of-mass of the pair $(\boldsymbol{p}_{1} + \boldsymbol{p}_{2} = 0)$, $\boldsymbol{r}$ is the relative coordinate with time difference correction, and $\Psi(\boldsymbol{r}, \boldsymbol{k})$ is the relative wave function in the two-body outgoing state with an asymptotic relative momentum $\boldsymbol{k}$.
In the case where we can ignore the time difference of the emission and the momentum dependence of the source, we integrate out the center-of-mass coordinate and obtain Eq.~(\ref{ignore}), where $S_{12}(r)$ is the normalized pair source function in the relative coordinate,given by the expression:
\begin{align}
	S_{12}(r) = \frac{1}{(4 \pi R^2)^{3/2}} \text{exp}(-\frac{r^2}{4R^2}),
\end{align}
where $R$ is the size parameter of the source.
Thus, two important factors of the correlation function are included in Eq.~(\ref{ignore}): the collision system, which is related to the source function $S_{12}(r)$, and the two-particle interaction, which is embedded in the relative wave function $\Psi(\boldsymbol{r}, \boldsymbol{k})$.

For a pair of non-identical particles, such as $p$-$\Omega$, assuming that only $S$-wave part of the wave function is modified by the two-particle interaction, $\Psi(\boldsymbol{r}, \boldsymbol{k})$ can be given by:
\begin{align}
	\Psi_{p \text{-} \Omega}(\boldsymbol{r}, \boldsymbol{k}) = \text{exp}(\text{i} \boldsymbol{k} \cdot \boldsymbol{r}) -j_0(kr) + \psi_{p \text{-} \Omega}(r,k),
\end{align}
where the spherical Bessel function $j_0(kr)$ represents the $S$-wave part of the non-interacting wave function, and $\psi_{p \text{-} \Omega}$ stands for the scattering wave function affected by the two-particle interaction.
Substituting the relative wave function $\Psi_{p \text{-} \Omega}(\boldsymbol{r}, \boldsymbol{k})$ into the KP formula yields the correlation function:
\begin{align}
	C_{p \text{-} \Omega}(k) = 1 + \int_{0}^{\infty} 4\pi r^2 \, \mathrm{d}r \, S_{12}(r) \, [ |\psi_{p \text{-} \Omega}(r,k)|^2 - |j_0(kr)|^2 ]. \label{Ck}
\end{align}

$\psi_{p \text{-} \Omega}(r,k)$ can be obtained by solving the Schr\"{o}dinger equation, and a similar approach has been utilized in the femtoscopic correlation analysis tool using the Schr\"{o}dinger equation~\cite{Mihaylov:2018rva}:
\begin{align}
	-\frac{\hbar ^{2}}{2\mu }\nabla ^{2}\psi_{p \text{-} \Omega}(r,k) +V(r) \psi_{p \text{-} \Omega}(r,k) = E \psi_{p \text{-} \Omega}(r,k)
\end{align}
where $\mu = m_p m_\Omega / (m_p + m_\Omega)$ is the reduced mass of the system.

Considering the case of the $S$-wave, the wave function can be separated into a radial term $R_k(r)$ and an angular term $Y_0^0(\theta, \phi)$ and expressed as:
\begin{align}
	\psi_{p \text{-} \Omega}(r,\theta, \phi) =  R_k(r) Y_0^0(\theta, \phi).
\end{align}

Considering the interaction between a proton and an $\Omega$ baryon, which includes both the strong interaction and the Coulomb interaction, the potential can be written as:
\begin{align}
	V(r) = V_{\text{Strong} }(r) + V_{\text{Coulomb}} (r),   \label{Coulomb}
\end{align}
where $V_{\text{Coulomb}}(r) = - \alpha \hbar c/r $, and $\alpha$ is the fine-structure constant.
The method to obtain the strong interaction potential $V_{\text{Strong} }(r)$ will be introduced in the next section.

Once the total interaction potential is determined, the radial Schr\"{o}dinger equation can be solved:
\begin{align}
	\frac{-\hbar^{2}}{2 \mu} \frac{\text{d}^{2}u_k(r)}{\text{d}r^{2}} + V(r) u_k(r)  = E u_k(r),   \label{eq}
\end{align}
where $E = \hbar^2 k^{2} / (2 \mu)$ and $u_k(r) = r R_k(r)$.
On this basis, the correlation function $C_{p \text{-} \Omega}(k)$ for given spin-parity quantum numbers can be calculated through Eq.~(\ref{Ck}).
The calculation of the correlation functions described above is based on obtaining the scattering wave functions by solving the Schr\"{o}dinger equation in coordinate space~\cite{Morita:2014kza,Ohnishi:2016elb,Hatsuda:2017uxk,Kamiya:2019uiw,Kamiya:2021hdb,Ohnishi:2021ger,Ogata:2021mbo}.
Additionally, the scattering wave functions can also be obtained by solving the Lippmann-Schwinger (Bethe-Salpeter) equation in momentum space~\cite{Haidenbauer:2018jvl,Haidenbauer:2020kwo,Liu:2022nec,Liu:2023uly,Liu:2024nac}.
Further details on correlation functions for various systems can be found in the references mentioned above.

Additionally, for the $S$-wave $p$-$\Omega$ dibaryon system, the possible spin-parity quantum numbers can be $J^P = 1^+$ and $2^+$, respectively.
Since the experimentally measured correlation function is spin-averaged, the theoretically obtained correlation function should also consider the average over systems with different quantum numbers:
\begin{align}
	C_{p \text{-} \Omega}(k) = \frac{3}{8}C_{p \text{-} \Omega}^{J = 1}(k) + \frac{5}{8} C_{p \text{-} \Omega}^{J = 2}(k).  \label{average}
\end{align}

\begin{figure}[htb]
	\centering
	\includegraphics[width=8cm]{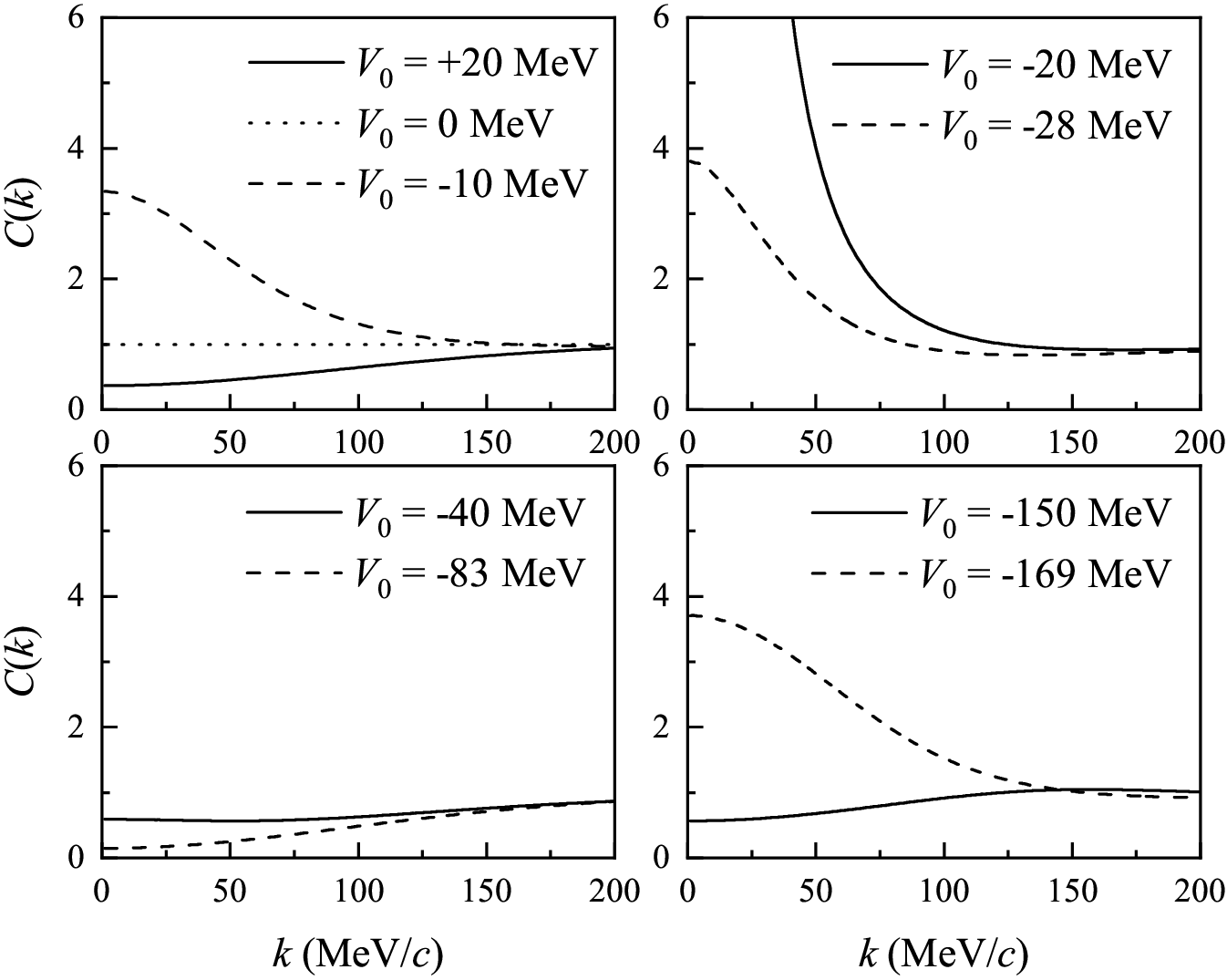}\
	\caption{The correlation functions for various squre potential models $V(r) = V_0 \, \theta(r_0 -r)$, where $r_0 = 2$ fm.}
	\label{figure 1}
\end{figure}

To understand the relationship between potentials and correlation functions, we here discuss the correlation functions for simplified square potential models.
We use the masses of $p$-$\Omega$ pair and the size parameter $R$ in source function from this work and the results are placed in four panels within Fig.~\ref{figure 1}.
First, for a repulsive potential ($V(r) = V_0 \, \theta(r_0 -r)$ with $V_0 = +20$ MeV and $r_0 = 2$ fm), the corresponding correlation function gradually increases from its initial position, approaching but remaining below unity.
Next, we set up square-well potentials of different depths.
The width of the square-well potentials remain $r_0 = 2$ fm, while the depth increases from 0 MeV to $-$10 MeV, $-$20 MeV, $-$28 MeV, $-$40 MeV, $-$83 MeV, $-$150 MeV, and $-$169 MeV.

When the potential depth is 0 MeV, meaning there is no interaction, the corresponding correlation function remains at unity.
As the attraction gradually strengthens, the position of the corresponding correlation function also rises ($V_0 = -10$ MeV).
When the depth of the square-well potential is $V_0 = -20$ MeV, the amplitude of the corresponding correlation function reaches its maximum value.
Subsequently, as the attraction gradually strengthens, the position of the correlation function will begin to decrease ($V_0 = -28$ MeV).
This downward trend continues until the corresponding correlation function drops below unity ($V_0 = -40$ MeV).
When the depth of the square-well potential is $V_0 = -83$ MeV, the amplitude of the corresponding correlation function reaches its minimum value.
A similar observation was also mentioned in Ref.~\cite{Liu:2023uly}, where the authors noted that correlation functions exhibit similar shapes for repulsive and strongly attractive potentials.
After reaching the minimum amplitude of the correlation function, as the attraction intensifies, the amplitude begins to rise again. 
This periodic-like variation persists.

This phenomenon is related to the periodic changes in the scattering length and effective range.
The correlation function is closely related to the scattering parameters~\cite{ExHIC:2017smd}, as seen in Lednicky-Lyuboshitz formula.
Therefore, as the attractive potential deepens, the scattering parameters change periodically with the appearance of new bound states, and the correlation function exhibits a periodic-like behavior.

\subsection{Gel'fand-Levitan-Marchenko method}

Obviously, to solve Eq.~(\ref{eq}), the two-body interaction potential $V(r)$ is essential.
The QDCSM is actually a treatment for the few-body problem, meaning that directly extracting a two-body interaction potential $V(r)$ from it is not straightforward, since the hadronization process is not fully complete.
Fortunately, the QDCSM can be employed to investigate the scattering process, which allows us to obtain the potential we need, as hadronization is fully completed in this approach.

The approach we adopted to extract the two-body effective potential $V(r)$ is the GLM method, a powerful tool in inverse scattering theory~\cite{Chadan1977}.
It provides a systematic approach to reconstruct an effective potential from the scattering data of a specific process, making it a classic example of an ``inverse problem''.
Thus, this method provides another avenue to understand the nature of the two-body interaction.
Furthermore, using the obtained potential, a series of studies can be conducted, such as calculating the spectrum of few-body systems~\cite{HALQCD:2014okw,HALQCD:2018qyu}, estimating production~\cite{Zhang:2020dma}, or investigating other experimental observables like the correlation functions~\cite{Morita:2016auo,Morita:2019rph} in this work.

The key equation of the GLM method used in this work is the Marchenko equation~\cite{Marchenko1955,Agranovich1963}, which can be written in the $S$-wave case as an integral equation:
\begin{align}
	K(r, r^{\prime}) + F(r, r^{\prime}) + \int_r^\infty K(r, s) F(s , r^{\prime}) \, \mathrm{d}s = 0.
\end{align}
Here, the kernel function $K(r, r^{\prime})$ is the solution to be determined, and $F(r, r^{\prime})$ is the inverse Fourier transform of the reflection coefficient, given by:
\begin{align}
	F(r, r^{\prime}) =& \frac{1}{2 \pi} \int_{-\infty}^{\infty} e^{\mathrm{i}kr}\left\{1-S(k)\right\} e^{\mathrm{i}kr^\prime} \mathrm{d} k \nonumber \\
	&+\sum_{i=1}^{n} M_{i} e^{-\kappa_i r} e^{-\kappa_i r^\prime}.
\end{align}
The partial-wave scattering matrix $S(k)$ is given by $S(k) = \mathrm{exp}(2 \mathrm{i} \delta(k))$, where $\delta(k)$ is the scattering phase shift, satisfying $k \cot \delta = -1/a_0 + 1/2 \, r_{\text{eff}} k^2$.
Here, $a_0$ and $r_{\text{eff}}$ represent the scattering length and the effective range, respectively, which are calculated in our previous work~\cite{Huang:2015yza}.
Additionally, $n$ is the number of bound states, $\kappa_i$ denotes the wavenumber of the $i$-th bound state, and $M_i$ is the norming constant.
Then, after solving Marchenko equation and obtaining $K(r,r^\prime)$, the potential can be reconstructed as:
\begin{equation}
	V(r)=-2\frac{\mathrm{d}}{\mathrm{d}r}K(r,r).
\end{equation}

There is one point we would like to emphasize.
Generally, when bound states exist, this method does not provide a fully determined potential, but instead results in a set of phase-effective potentials~\cite{Sofianos:1990rb}.
However, if one fixes all the $M_i$ in a unique way, such as by calculating from the Jost solution, the obtained potential will be unique for further calculations~\cite{Massen:1999vi,Newton2002}.
By using this method, preparation for calculating correlation functions has been done.
In addition, calculations related to hypernuclei can also be performed based on the obtained hadron-hadron potentials~\cite{STAR:2010gyg,Garcilazo:2019igo,Hiyama:2019kpw,Zhang:2021vsf,Chen:2023mel,STAR:2022fnj,Ma:2023,STAR:2023fbc,Hiyama:2022jqh}.
For a more comprehensive discussion on the GLM method, one can refer to Refs.~\cite{Chadan1977,Marchenko1955,Agranovich1963,Sofianos:1990rb,Massen:1999vi,Newton2002,Jade:1996am,Meoto:2019jky,Khokhlov:2021afy,Khokhlov:2022uie}.

\section{The results and discussions}
\label{sec3}

\begin{figure}[htb]
	\centering
	\includegraphics[width=7cm]{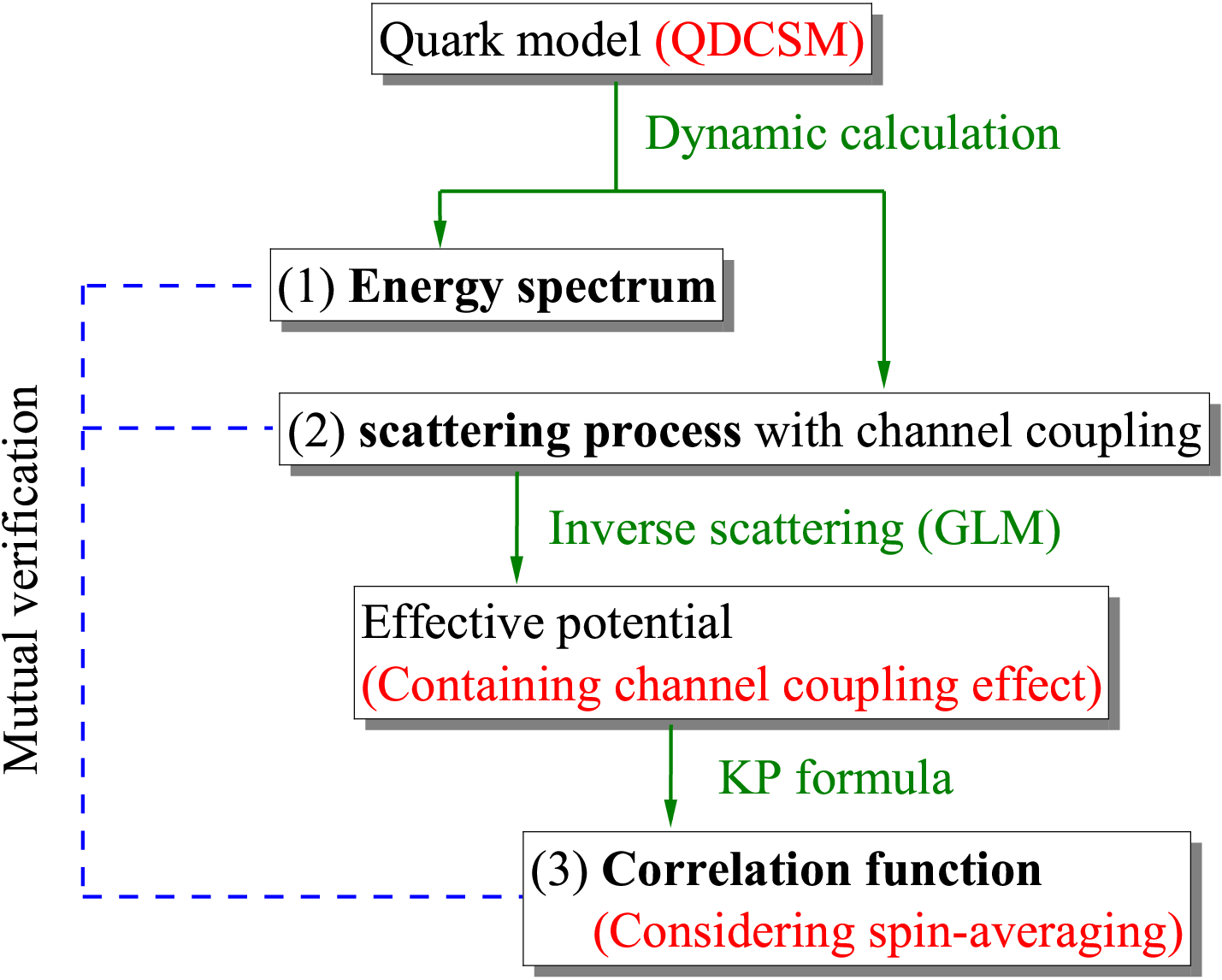}\
	\caption{The simplified diagram of the process in this work.}
	\label{figure 2}
\end{figure}

The simplified diagram of the process in this work is shown in Fig.~\ref{figure 2}.
The $S$-wave $p$-$\Omega$ dibaryon systems are studied using the QDCSM, and an extended Kohn-Hulth\'{e}n-Kato (KHK) variational method is employed to investigate the $p$-$\Omega$ scattering processes.
Both single channel calculation and coupled channel calculation are carried out. 
For the $J^P = 1^+$ system, the coupled channels are $p$-$\Omega$, $\Xi$-$\Sigma$, $\Xi$-$\Lambda $, $\Xi^*$-$\Lambda$, $\Xi^*$-$\Sigma$, $\Xi$-$\Sigma^*$, and $\Xi^*$-$\Sigma^*$.
For the $J^P = 2^+$ system, the coupled channels are $p$-$\Omega$, $\Xi^*$-$\Lambda$, $\Xi^*$-$\Sigma$, $\Xi$-$\Sigma^*$, and $\Xi^*$-$\Sigma^*$.
Since this work mainly focuses on the correlation functions, the details of the QDCSM can be seen in Ref.~\cite{Huang:2015yza,Liu:2022vyy}, and the details of the KHK variational method and scattering process calculations can be found in Appendix and Refs.~\cite{Yan:2024usf,Kamimura:1977okl}.

For the energy spectrum, on the basis of our result~\cite{Huang:2015yza}, the $S$-wave $p$-$\Omega$ systems with $J^P = 1^+$ and $2^+$ are both unbound in the single channel calculation.
However, after channel coupling, a weakly bound state with a binding energy of 5 MeV is formed in the $J^P = 2^+$ $p$-$\Omega$ system, while the $J^P = 1^+$ $p$-$\Omega$ system is still unbound.
Early theoretical works mainly focused on confirming the existence of the $J^P = 2^+$ $p$-$\Omega$ bound state~\cite{Goldman:1987ma,Oka:1988yq,Silvestre-Brac:1992xsl}.
Most recent quark model calculations support that the $p$-$\Omega$ system with $J^P = 2^+$ forms a weakly bound state, which can be explained by the limited interaction between $p$ and $\Omega$ because they do not have common flavor quarks~\cite{Huang:2015yza,Dai:2007gc,Sekihara:2018tsb,Sekihara:2023ihc}. 
Our work also proved that if the difference of quark favors between $u$ and $s$ quarks is ignored and the extent of interaction is artificially expanded, the binding energy of the $p$-$\Omega$ system can be deepened~\cite{Huang:2015yza}.
The HAL QCD collaboration reported two lattice QCD simulations on the $p$-$\Omega$ system with $J^P = 2^+$~\cite{HALQCD:2014okw,HALQCD:2018qyu}. 
The conclusion of a weakly bound state in the second simulation is more reliable since it is obtained under nearly physical quark masses condition.

\begin{figure}[htb]
	\centering
	\includegraphics[width=8cm]{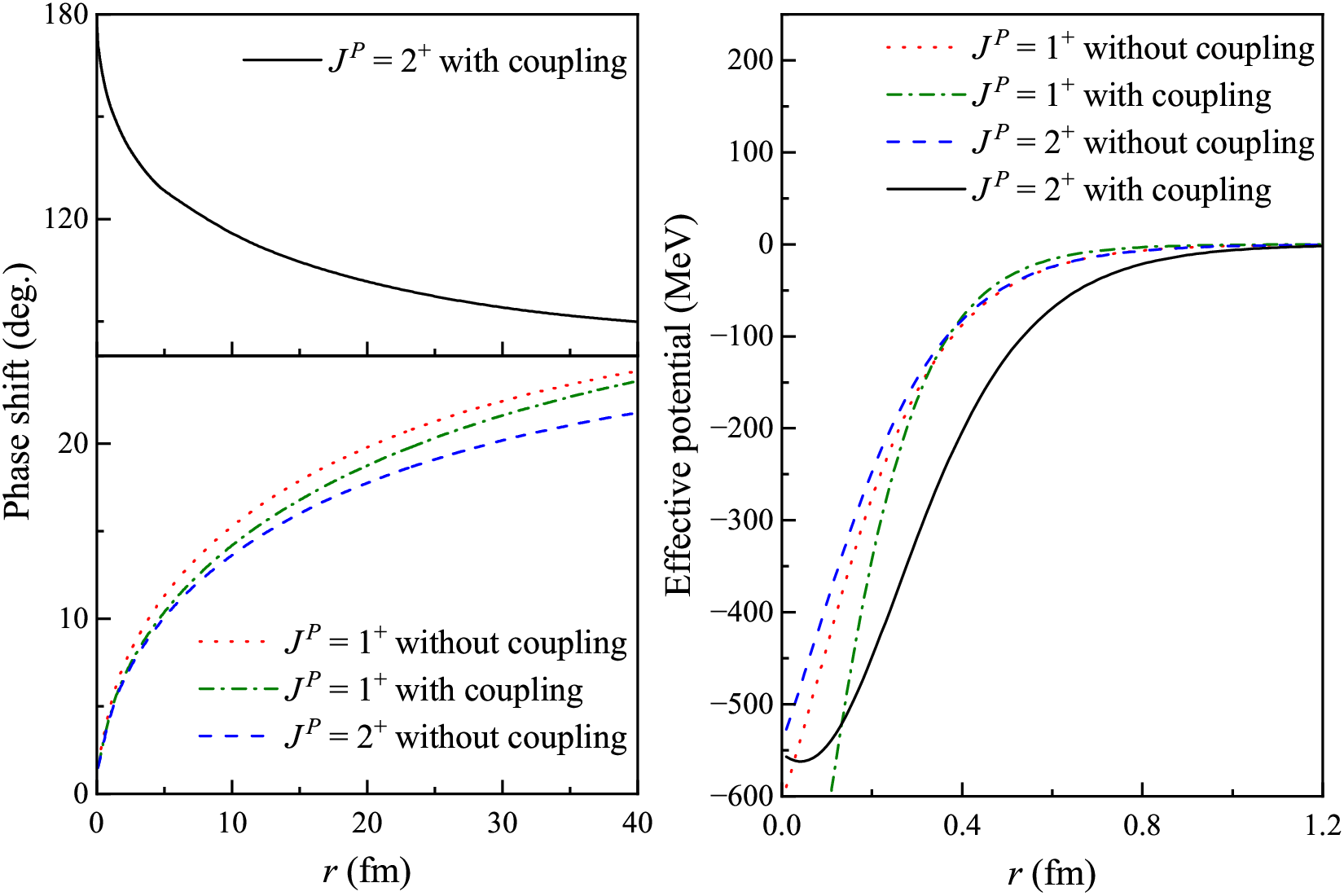}\
	\caption{The scattering phase shifts and effective potentials of the $p$-$\Omega$ dibaryon systems with and without channel coupling.}
	\label{figure 3}
\end{figure}

For the scattering process, both the $p$-$\Omega$ systems with and without channel coupling are studied.
The scattering phase shifts of the $p$-$\Omega$ systems are shown in the left panel of Fig.~\ref{figure 3}.
It can be seen that when channel coupling is not considered, the scattering phase shifts of the $p$-$\Omega$ with $J^P = 1^+$ (dotted red line) and $2^+$ (dashed blue line) in single channels are very close, indicating that the $p$-$\Omega$ interactions in the two single channels are similar.
After accounting for the channel coupling, as the incident energy approaches 0 MeV, the phase shift of the $p$-$\Omega$ with $J^P = 2^+$ (solid black line) tends to $180^\circ$ , confirming the existence of a $p$-$\Omega$ bound state.
For the $p$-$\Omega$ system with $J^P = 1^+$, the phase shift after channel coupling also changes compared to that calculated for a single channel.
Since the $J^P = 2^+$ $p$-$\Omega$ forms a bound state under the influence of other channels, the effect of channel coupling on the phase shift of the $J^P = 2^+$ $p$-$\Omega$ is significant, and it similarly affects the effective potential.
For the $J^P = 1^+$ $p$-$\Omega$, no bound state is formed due to the influence of the inelastic channels with lower thresholds, such as $\Xi$-$\Lambda$. 
Therefore, the impact of channel coupling on the phase shift of $J^P = 1^+$ $p$-$\Omega$ is not as significant as that of $J^P = 2^+$ $p$-$\Omega$.
We summarize the properties and scattering parameters of the $p$-$\Omega$ systems in Table~\ref{result}.
It is worth noting that these results have not taken the Coulomb interaction into consideration.

\begin{table} [htb]
	\caption{\label{hadrons} The binding energy $E_B$, scattering length $a_0$, and effective range $r_\text{eff}$ of the $p$-$\Omega$ system with $J^P = 1^+$ and $2^+$ with and without channel coupling.}
	\begin{tabular}{c c c c c}
		\hline \hline
		$p$-$\Omega$     & ~~$J^P$ & $E_B$ & $a_0$ (fm) & $r_\text{eff}$ (fm)  \\
		\hline
		without         & $1^+$ & unbound & $-0.53$ & 0.96  \\
		channel coupling & $2^+$ & unbound & $-0.47$ & 1.06  \\ 
		\hline
		with   & $1^+$ & unbound & $-0.48$ & 0.74  \\
		channel coupling & $2^+$ & 5 MeV   & 2.80    & 0.58  \\
		\hline\hline
		\label{result}
	\end{tabular}
\end{table}

In Ref.~\cite{HALQCD:2018qyu}, the lattice QCD simulation for the $J^P = 2^+$ $p$-$\Omega$ single channel yielded a binding energy of 1.54 MeV and a scattering length of 5.30 fm.
Our results show a binding energy and scattering length of 5 MeV and 2.8 fm, respectively, after considering channel coupling.
Since the binding energy in our work is larger, the scattering length is correspondingly smaller and positive.
For the $p$-$\Omega$ with $J^P = 1^+$, although it is unbound, its scattering phase shift remains positive, indicating the presence of an attractive potential.
Based on the scattering parameters, the effective potentials of the $p$-$\Omega$ systems can be calculated using the GLM method.
It is important to note that the inverse scattering problem cannot be uniquely solved in this context. 
The effective potential obtained through the GLM method represents one solution within this framework, accurately reproducing the binding energy, scattering length, and effective range.
Additionally, using the scattering data from the lattice simulation, we have reproduced the corresponding correlation function.
This ensures the reliability of the effective potentials obtained via the GLM method in calculating correlation functions.

The corresponding effective potentials are shown in the right panel of Fig.~\ref{figure 3}.
By inverting the $p$-$\Omega$ phase shifts obtained with channel coupling, the coupling to other channels is renormalized into an effective single channel $p$-$\Omega$ potential.
It is worth noting that the effect of channel coupling on the correlation function is incorporated in this process, relating its calculation to the traditional scattering process. 
This approach differs from that taken in Refs.~\cite{Liu:2022nec,Kamiya:2019uiw} regarding channel coupling.
A similar strategy to this work was used in Ref.~\cite{Hiyama:2019kpw}, where E. Hiyama \textit{et al.} introduced an effective single-channel $\Xi$-$N$ potential. 
In this potential, the coupling to $\Lambda$-$\Lambda$ in $^{11} S_{0}$ is renormalized into a single-range Gaussian form $U_{2} \cdot \exp \left(-(r / \gamma)^{2}\right)$ with $\gamma=1.0$ fm and $U_{2}$ $(<0)$ chosen to reproduce the $\Xi$-$N$ phase shifts obtained with channel coupling.
The difference from Ref.~\cite{Hiyama:2019kpw} is that we use a quark model to calculate the coupled channel scattering process and employ the GLM method to solve the inverse scattering problem.
Moreover, the $\Xi$-$N$ potential obtained in Ref.~\cite{Hiyama:2019kpw} has been further folded into a $\Xi$-$\alpha$ potential~\cite{Hiyama:2022jqh} and used to calculate the $\Xi$-$\alpha$ correlation function~\cite{Kamiya:2024diw,Jinno:2024rxw}.

\begin{figure*}[htb]
	\centering
	\includegraphics[width=16cm]{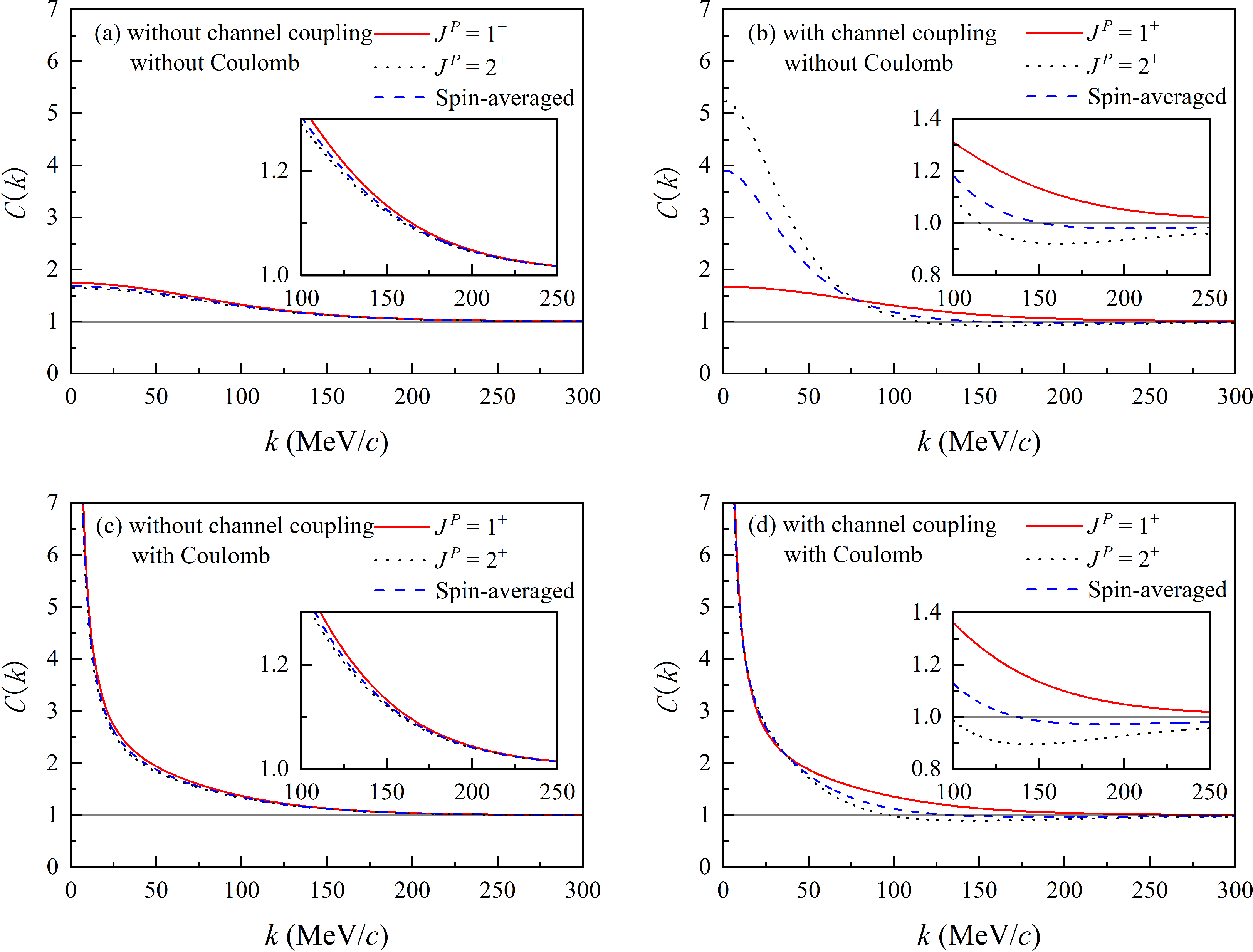}\
	\caption{The correlation functions for the $p$-$\Omega$ system are presented for the $J^P = 1^+$, $J^P = 2^+$, and spin-averaged cases. 
		The results in panel (a) are calculated without channel coupling and without considering the Coulomb interaction. 
		The results in panel (b) are calculated with channel coupling, but without the Coulomb interaction included. 
		The results in panel (c) are calculated without channel coupling, but with considering the Coulomb interaction. 
		The results in panel (d) are calculated with channel coupling and with the Coulomb interaction included.}
	\label{figure 4}
\end{figure*}

Using the effective potentials obtained above, we can further calculate the $p$-$\Omega$ correlation functions.
The value of the size parameter $R =0.95 \pm 0.06$ fm  of the source function $S_{12}(r)$ in the KP formula is taken from the experimental measurement~\cite{ALICE:2020mfd} and determined via an independent analysis of $p$-$p$ correlations~\cite{ALICE:2020ibs}.
The $p$-$\Omega$ correlation functions under different conditions are shown in Fig.~\ref{figure 4}.
Here, we aim to discuss the effect of the Coulomb interaction as well as spin-averaging.
Therefore we only take the central value the size parameter $R$ to calculate the $p$-$\Omega$ correlation functions in this part.
The Error of the size parameter will be considered in the next part when comparing with experimental measurements.
In addition, the Coulomb interaction and spin-averaging are taken into consideration according to Eq.~(\ref{Coulomb}) and Eq.~(\ref{average}), respectively.

In Fig.~\ref{figure 4}, panels (a) and (b) show the results before and after channel coupling without considering the Coulomb interaction, while panels (c) and (d) show the results before and after channel coupling with the Coulomb interaction included.
It is evident that the attractive Coulomb interaction, despite its relatively weak strength as a long-range force, significantly enhances the amplitude of the correlation functions under various conditions.
Therefore, when calculating correlation functions for two charged particles, the impact of the Coulomb interaction must be taken into account.

In both panels (a) and (c), since the attractive interactions in these systems are too weak to form bound states, the corresponding correlation functions remain above unity.
In panels (b) and (d), we can see that the existence of the $J^P = 2^+$ bound state leads to the depletion of the correlation function around $k$ = 150 MeV/$c$ (dotted black lines).
This depletion, caused by a bound state, is consistent with the lattice QCD simulations~\cite{HALQCD:2018qyu,Morita:2019rph}.
After considering spin-averaging, i.e., after the weighted summation of the correlation functions for $J^P = 1^+$ and $J^P = 2^+$ $p$-$\Omega$ according to the spin quantum number, the total correlation function (dashed blue lines) lies between the original two.
In the enlarged images in panels (b) and (d), it can be seen that the depletion of the correlation function becomes less obvious due to spin-averaging.

\begin{figure}[htb]
	\centering
	\includegraphics[width=8cm]{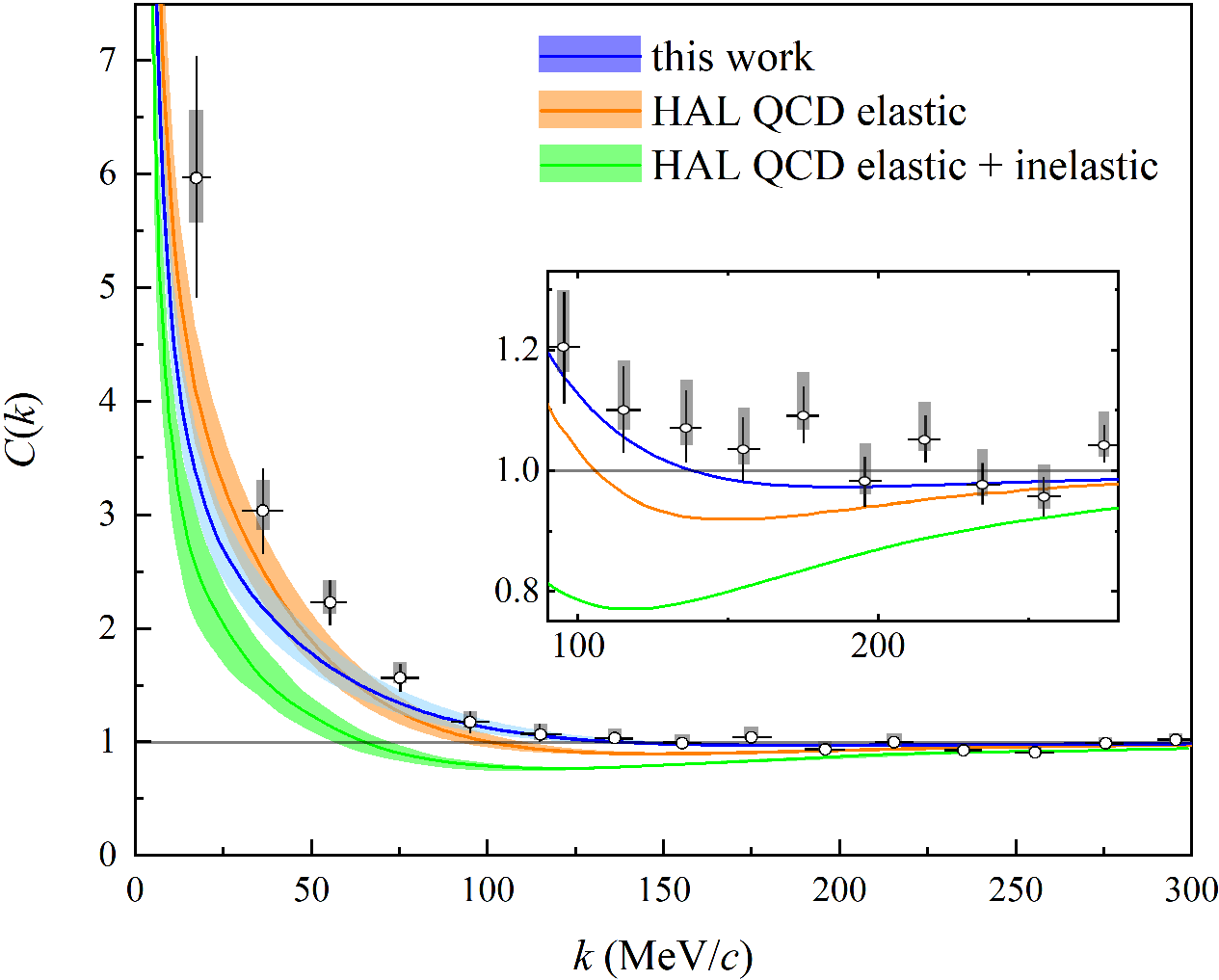}\
	\caption{
		The correlation functions of the $p$-$\Omega$ system.
		The blue band represents the result from our model (QDCSM).
		The orange band represents the result from lattice QCD considering only the $J^P = 2^+$ elastic contribution and the Coulomb interaction (HAL QCD)~\cite{HALQCD:2018qyu}.
		The green band represents the result from Ref.~\cite{Morita:2019rph}, based on the lattice QCD simulation for $J^P = 2^+$, further assuming complete absorption of $p$-$\Omega$ pairs with $J^P = 1^+$ into octet-octet states (HAL QCD + complete absorption).
		The pink band represents the result from our model with further assumption of the same complete absorption with Ref.~\cite{Morita:2019rph} (QDCSM + complete absorption).
		The black vertical bars and grey boxes represent the statistical and systematic uncertainties of the experimental data~\cite{ALICE:2020mfd}.}
	\label{figure 5}
\end{figure}

After accounting for the $p$-$\Omega$ strong interaction, channel coupling, Coulomb interaction, spin-averaging, and the error in the size parameter $R$ of the source function, we can compare the computed correlation function with the latest experimental measurement of the $p$-$\Omega$ correlation~\cite{ALICE:2020mfd}.
The theoretical results and experimental measurements are shown in Fig.~\ref{figure 5}.
The behavior of the correlation functions obtained under different conditions is similar in the low-energy region ($k$ = 0--15 MeV/$c$).
This is due to the inclusion of an attractive Coulomb potential and a $J^P = 2^+$ bound state.
Theoretically obtained correlation functions in this region appear to be lower than the experimental measurements.
However, the experimental measurements in this region have relatively large uncertainties.
More precise measurements would aid further analysis and understanding of the reasons behind this discrepancy.

Before the introduction of complete absorption effect, in the $k$ = 15--60 MeV/$c$ region, the result of the QDCSM is close to that of the HAL QCD but still lower.
This is because the binding energy we obtained for the $J^P = 2^+$ bound state is larger, resulting in a stronger attractive potential.
Our calculations show that, when the $p$-$\Omega$ system forms a weakly bound state, the amplitude of the correlation function decreases as the attractive potential strengthens.
Thus, the contribution from the $J^P = 2^+$ bound state causes our result to be lower than that of HAL QCD.

In the $k$ = 60--200 MeV/$c$ region, one of the questions raised after the ALICE collaboration's measurements was why the depletion of the correlation function, seen in the calculations around $k$ = 150 MeV/$c$ due to the presence of a $p$-$\Omega$ bound state, is not observed in the measured data.
According to our calculation, this can be explained by the contribution from the $J^P = 1^+$ $p$-$\Omega$.
Since the potential of the $J^P = 1^+$ $p$-$\Omega$ is attractive and no bound state is formed, the $J^P = 1^+$ $p$-$\Omega$ correlation remains above unity.
The Coulomb interaction further enhances this phenomenon.
After spin-averaging, the depletion caused by $J^P = 2^+$ bound state becomes less pronounced.
In the $k > 200$ MeV/$c$ region, the correlation functions asymptotically approach unity.

A sign of the $p$-$\Omega$ correlation function's subtle sub-unity part can also be seen in experimental measurements.
However, after considering the statistical and systematic uncertainties of the experimental data, the experimental conclusions regarding the sub-unity part are not sufficiently certain. 
Future experimental measurements could focus on examining this sub-unity part of the $p$-$\Omega$ correlation function.
In Fig.~\ref{figure 5}, the error represented by the band arises from the uncertainty in the size parameter $R$ of the source function.
It can be seen that the value of the size parameter $R$ has a significant impact on the correlation functions.
This issue has already been addressed in theoretical studies~\cite{Morita:2016auo,Mihaylov:2018rva}, and we also plan to investigate it further in the future.

In Ref.~\cite{Morita:2019rph}, K. Morita \textit{et al.} introduced the absorption effect based on the HAL QCD results, assuming that the $J^P = 1^+$ $p$-$\Omega$ wave function is completely absorbed into octet-octet states.
In our calculations, we first account for the effect of other physical channels by solving the inverse scattering problem of the coupled channels.
We also consider the introduction of absorption effect in the same way as in Ref.~\cite{Morita:2019rph} and the corresponding results are shown in Fig.~\ref{figure 5}.
For the $J^P = 1^+$ $p$-$\Omega$, the real and imaginary parts of the potential are given by:
\begin{align}
	&V^{J=1}_\text{Real} (r) = V_\text{Strong}(r) + V_\text{Coulomb}(r), \nonumber \\
	&V^{J=1}_\text{Imaginary} (r) = - \text{i} V_0 \theta(r_0 -r). \label{im}
\end{align}
In the case where the $J^P = 1^+$ $p$-$\Omega$ is completely absorbed, the correlation functions are significantly reduced compared to those without this effect, as the correlation functions with $J^P = 1^+$ are much lower than unity.
Moreover, since $V_0$ in Eq.~(\ref{im}) tends to $+ \infty$ MeV, the correlation function with $J^P = 1^+$ is dominated by the absorption effect, rendering the strong interaction potential of the $J^P = 1^+$ $p$-$\Omega$ almost ineffective.
In this case, it is difficult to explain why the depletion of the correlation function, which is attributed to the $J^P = 2^+$ $p$-$\Omega$ bound state, is not observed experimentally.

\begin{figure}[htb]
	\centering
	\includegraphics[width=8cm]{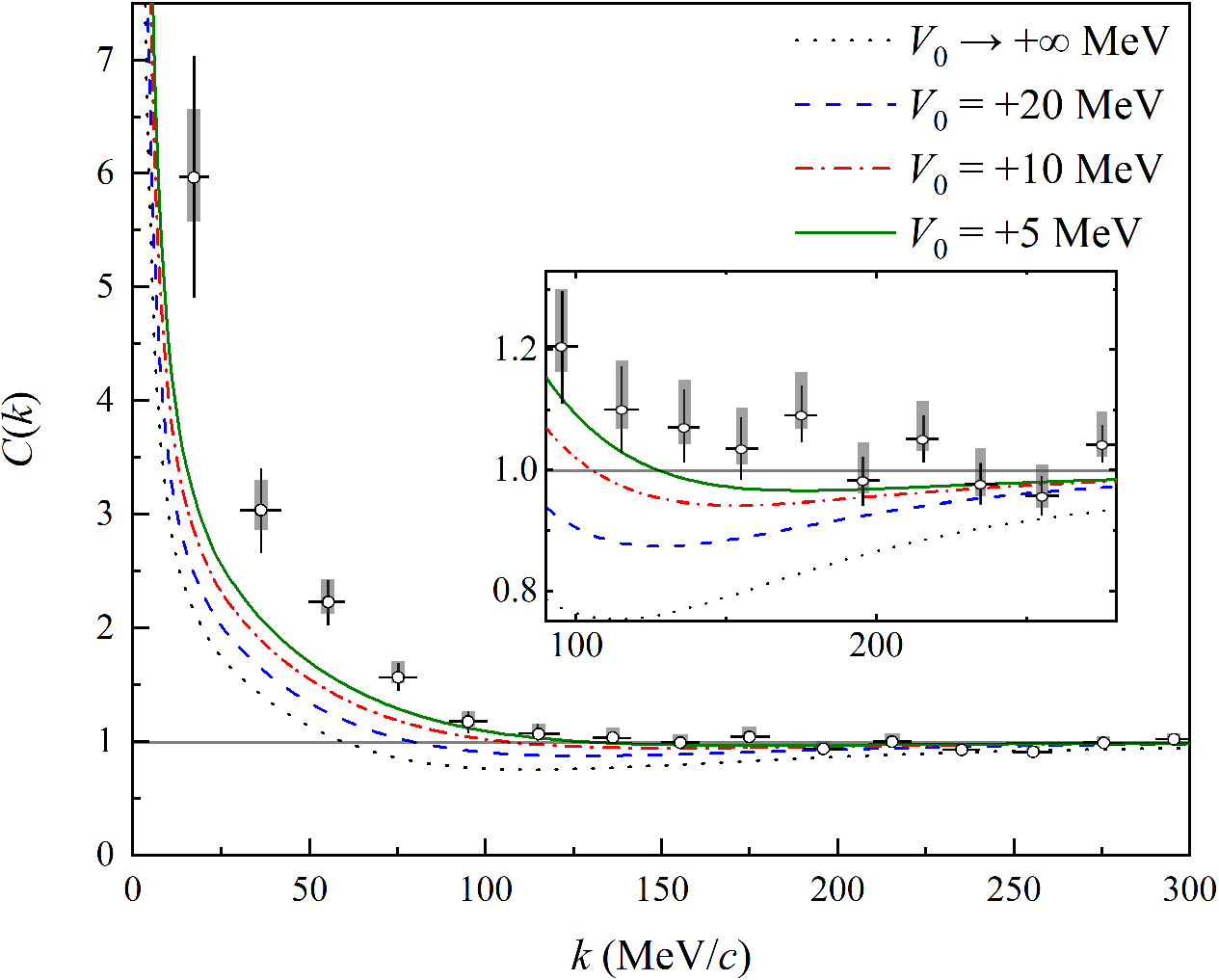}\
	\caption{
		The $p$-$\Omega$ correlation functions after further assuming different degrees of absorption effect.
		The imaginary part of the $p$-$\Omega$ potential with $J^P = 1^+$ is considered as $V(r) = - \text{i} V_0 \theta(r_0 -r) $.}
	\label{figure 6}
\end{figure}

However, the assumption that the $J^P = 1^+$ $p$-$\Omega$ is completely absorbed represents a theoretical extreme case. 
In reality, a portion of the $J^P = 1^+$ $p$-$\Omega$ may be converted into octet-octet pairs via inelastic scattering, while another portion may be converted into $p$-$\Omega$ pairs via elastic scattering.
In this case, both the absorption effect and the $J^P = 1^+$ $p$-$\Omega$ strong interaction can influence the correlation functions.
We adjust the effect of the absorption by varying the value of $V_0$ in Eq.~(\ref{im}), and the corresponding correlation functions are shown in Fig.~\ref{figure 6}.
To observe the changes of correlation functions, we only take the central value of the size parameter $R$ in the calculation.
As the value of $V_0$ decreases from infinity, the absorption effect gradually weakens, and the theoretical results converge towards the experimental measurements.
However, it is currently difficult to determine the appropriate value of $V_0$ to represent the actual physical situation.
Experimental measurements of the $\Lambda$-$\Xi$ and $\Sigma$-$\Xi$ correlation functions in the future will help address this issue.

In general, correlation functions provide us with a pathway to study hadron-hadron interactions and exotic hadronic states.
For instance, Z. W. Liu \textit{et al.} have already employed correlation functions to investigate $Z_c(3900)$ and $Z_{cs}^*(3985)$~\cite{Liu:2024nac}.
And N.~Ikeno  \textit{et al.} have investigated the inverse problem of extracting information from the correlation functions to study $D^*_{s0}(2317)$~\cite{Ikeno:2023ojl}.
The study of correlation functions and exotic hadronic states will require collaborative efforts between theory and experiment in the future.

\section{Summary}

In this work, the $p$-$\Omega$ interactions are studied by calculating the $p$-$\Omega$ correlation functions.
The $p$-$\Omega$ strong interaction, channel coupling, the Coulomb interaction, and spin-averaging are considered.
The coupling to other channels is renormalized into an effective single channel $p$-$\Omega$ potential by inverting the $p$-$\Omega$ phase shifts obtained with channel coupling.

Based on the current results, the conclusion can be drawn as follows: 
(1) The depletion of the correlation function attributed to the $J^P = 2^+$ $p$-$\Omega$ bound state may be less pronounced due to the contribution of the attractive $J^P = 1^+$ $p$-$\Omega$ component in spin-averaging.
This can explain that why the Alice collaboration did not observe the obvious depletion caused by the possible $p$-$\Omega$ bound state, which is predicted by our model calculation and lattice QCD simulation. 
(2) The impact of the Coulomb interaction cannot be neglected when calculating correlation functions for two charged particles. 
In particular, the correlation function in the low energy region is obviously affected by the Coulomb interaction.
(3) We suggest that future experimental measurements focusing on examining the sub-unity part of the correlation function and performing more measurements with adjusted source sizes will help us further understand the $p$-$\Omega$ interaction.
(4) Introducing the absorption effect can lead to a noticeable decrease in the $p$-$\Omega$ correlation function with $J^P = 1^+$. 
Future experiments can measure the correlation functions of $\Lambda$-$\Xi$ and $\Sigma$-$\Xi$ to test the coupling strength between these systems and the $p$-$\Omega$ system.

We have systematically studied the $p$-$\Omega$ system from the perspective of the quark model, focusing on the energy spectrum, scattering phase shifts, and correlation functions.
A consistent conclusion is obtained from all three aspects of the investigation, and the existence of the $p$-$\Omega$ state is supported by our calculations.
This conclusion is consistent with the lattice QCD study as well.
These results demonstrate that the quark model is an effective method with predictive power for studying dibaryon systems.
Future experimental measurements of the $p$-$\Omega$ system may help discover a new dibaryon in addition to the deuteron and $d^*$. Furthermore, studying the $\Lambda$-$\Xi$ and $\Sigma$-$\Xi$ correlations can provide additional theoretical insights to assist in the search for the $p$-$\Omega$ state, which is part of our future work.

\acknowledgments{This work is supported partly by the National Natural Science Foundation of China under Contracts Nos. 11675080, 11775118, 12305087, 11535005 and 11865019. Y. Y. is supported by the Postgraduate Research and Practice Innovation Program of Jiangsu Province under Grant No. KYCX23\underline{~}1675 and the Doctoral Dissertation Topic Funding Program under Grant No. YXXT23-027. Q. H. is supported by the Start-up Funds of Nanjing Normal University under Grant No. 184080H201B20.}	

\setcounter{equation}{0}
\renewcommand\theequation{A\arabic{equation}}

\section*{Appendix A: Coupled channel scattering phase shift calculation}

Experimental scattering processes are often influenced by channel coupling, so it is necessary to consider multi-channel coupling.
For two-particle scattering, we assume the total wave function is $\Psi^{(c)}$, which satisfies the Schr"{o}dinger equation:
\begin{align}
	(H-E) \Psi^{(c)}=0, \quad c=\alpha, \beta.
\end{align}
Here, $\Psi^{(\alpha)}$ ($\Psi^{(\beta)}$) represents the wave function of the incident channel $\alpha$ ($\beta$), including all outgoing channels.
We make an approximation by omitting the wave function of the decay channel involving three or more bodies.
Under this approximation, the total wave function is written as:
\begin{align}
	\Psi^{(c)}=\sum_\gamma \mathcal{A}_\gamma\left[\Phi_\gamma(\hat{\xi}) \chi_\gamma^{(c)}\left(R_\gamma\right)\right]+\Omega^{(c)} \cdot(c=\alpha, \beta).
\end{align}
Here, $\gamma$ represents all two-body channels, and $\Omega^{(c)}$ represents the residual decay amplitude not included in the previous term, which can generally be omitted.
The asymptotic behavior of $\chi_\gamma^{(c)}(R_\gamma)$ is given by:
\begin{align}
	\chi_\gamma^{(c)}\left(R_\gamma\right)=\chi_\gamma^{(-)}\left(k_\gamma, R_\gamma\right) \delta_{\gamma c}+S_{\gamma c} \chi_\gamma^{(+)}\left(k_\gamma, R_\gamma\right),  R_\gamma>R_\gamma^{(c)},
\end{align}
where $S_{\gamma c}$ represents the matrix of $c \rightarrow \gamma$ $S$, and $\chi_\gamma^{( \pm)}$ satisfies:

i) for open channel $\left(E_\alpha > 0\right)$:
\begin{align}
	\chi_\alpha^{( \pm)}\left(k_\alpha, R_\alpha\right)=\frac{1}{\sqrt{v_\alpha}} h_{L_\alpha}^{( \pm)}\left(k_\alpha, R_\alpha\right), \quad R_\alpha>R_\alpha^C.
\end{align}

ii) for a closed channel $\left(E_\alpha < 0\right)$:
\begin{align}
	\chi_\alpha^{( \pm)}\left(k_\alpha, R_\alpha\right)= W_{L_\alpha}^{( \pm)}\left(k_\alpha, R_\alpha\right), \quad R_\alpha>R_\alpha^C.
\end{align}
Where $v_\alpha$ is the relative velocity, $v_\alpha = \hbar k_\alpha / \mu_\alpha$, and a discussion on $W_{L_\alpha}^{( \pm)}\left(k_\alpha, R_\alpha\right)$ can be seen in Ref.~\cite{Kamimura:1977okl}.

Similar to single channel calculation, we first introduce a trial wave function $\Psi_{t}^{(c)}$:
\begin{equation}
	\Psi_{t}^{(c)} = \sum_{\gamma} A_{\gamma} \left[ \Phi_{\gamma}(\xi) \chi_{\gamma,t}^{(c)}(R_{\gamma}) \right] + \sum_{\nu} b_{\nu}^{(c)} \Omega_{\nu,t}, \quad (c = \alpha, \beta),
\end{equation}
where $\gamma$ represents all two-body channels, and $\Omega_{\nu,t}$ represents the residual term that is not included in the previous part. $\chi_{\gamma,t}^{(c)}(R_{\gamma})$ is the relative motion orbital trial wave function, whose asymptotic behavior is:
\begin{equation}
	\chi_{\gamma,t}^{(c)}(R_{\gamma}) = \chi_{\gamma}^{(-)}(k_{\gamma}, R_{\gamma}) \delta_{\gamma c} + S_{\gamma,t}^{c} \chi_{\gamma}^{(+)}(k_{\gamma}, R_{\gamma}), R_{\gamma} > R_{\gamma}^{(c)}.
\end{equation}

Here, we expand $\chi_{\gamma,t}^{(c)}(R_{\gamma})$ in terms of a series of known wave functions:
\begin{equation}
	\chi_{\gamma,t}^{(c)}(R_{\gamma}) = \sum_{i=0}^{n_{\gamma}} C_{\gamma i}^{(c)} \chi_{\gamma i}(R_{\gamma}), \quad (c = \alpha, \beta),
\end{equation}
where
\begin{equation}
	\chi_{\gamma i}(R_{\gamma}) = \begin{cases} 
		p_{\gamma i} \chi_{\gamma i}^{(\text{in})}(R_{\gamma}), & R_{\gamma} < R_{\gamma}^{c}, \\
		\chi_{\gamma}^{(-)}(k_{\gamma}, R_{\gamma}) + s_{\gamma i} \chi_{\gamma}^{(+)}(k_{\gamma}, R_{\gamma}), & R_{\gamma} > R_{\gamma}^{c}.
	\end{cases}
\end{equation}
Since the wave function $\chi_{\gamma i}(R_{\gamma})$ is continuous, it must be continuous at $R = R_{\gamma}^{c}$:
\begin{equation}
	p_{\gamma i} \chi_{\gamma i}^{(\text{in})}(R_{\gamma}^{c}) = \chi_{\gamma}^{(-)}(k_{\gamma}, R_{\gamma}^{c}) + s_{\gamma i} \chi_{\gamma}^{(+)}(k_{\gamma}, R_{\gamma}^{c}).
\end{equation}
The derivative of the wave function is also continuous:
\begin{align}
	&p_{\gamma i} \frac{d}{dR_{\gamma}} \chi_{\gamma i}^{(\text{in})}(R_{\gamma})\bigg|_{R_{\gamma}=R_{\gamma}^{c}}   \nonumber \\
	&= \frac{d}{dR_{\gamma}} \left[ \chi_{\gamma}^{(-)}(k_{\gamma}, R_{\gamma}) - s_{\gamma i} \chi_{\gamma}^{(+)}(k_{\gamma}, R_{\gamma}) \right]_{R_{\gamma}=R_{\gamma}^{c}}
\end{align}
Using these two conditions, we can solve for the unknowns \(p_{\gamma i}\) and \(s_{\gamma i}\). 
The specific method for open channels is similar to the previous calculation for closed channels. 
For further details on open channels, please see Ref.~\cite{Kamimura:1977okl}.

After solving for \(p_{\gamma i}\) and \(s_{\gamma i}\), we can obtain \(\chi_{\gamma i}(R_{\gamma})\), and by determining the expansion coefficients \(C_{\gamma i}^{(c)}\), we can obtain \(\chi_{\gamma,t}^{(c)}(R_{\gamma})\).
We first need to solve for \(C_{\gamma i}^{(c)}\) below.

From Eqs.~(a9), (a10), and (a11), we obtain:
\begin{equation}
	\sum_{i=0}^{n_{\gamma}} C_{\gamma i}^{(c)} = \delta_{\gamma c} \quad (c = \alpha, \beta),
\end{equation}
and
\begin{equation}
	\sum_{i=0}^{n_{\gamma}} C_{\gamma i}^{(c)} s_{\gamma i} = S_{\gamma, t}^{c}. \quad (c = \alpha, \beta)  \label{a13}.
\end{equation}
Following the same method as for the single-channel calculation, we finally obtain two sets of linear equations:
\begin{align}
	\sum_{\delta} \sum_{j=1}^{n_{\delta}} L_{\gamma i, \delta j} C_{\delta j}^{(c)}+\sum_{\mu} L_{\gamma i, \mu} b_{\mu}^{(c)}=M_{\gamma i}^{(c)}, \quad(c=\alpha, \beta),   \label{a14} \\ 
	\sum_{\delta} \sum_{j=1}^{n_{\delta}} L_{\nu, \delta_{j}} C_{\delta_{j}}^{(c)}+\sum_{\mu} L_{\nu, \mu} b_{\mu}^{(c)}=M_{\nu}^{(c)}, \quad(c=\alpha, \beta).  \label{a15}
\end{align}
Eq.~(\ref{a14}) is valid for all $\gamma_i$, and $i = 1 \sim n_\gamma$. 
Eq.~(\ref{a15}) is valid for all $\nu$.
\begin{align}
	K_{\gamma i, \delta j} & =\int A_{\gamma}\left[\phi_{\gamma}^{\dagger} \chi_{\gamma i}\right](H-E) A_{\delta}\left[\phi_{\delta} \chi_{\delta j}\right] d \tau,  \nonumber \\
	K_{\nu, \delta j} & =\int \tilde{\Omega}_{\bar{\nu}, t}^{\dagger}(H-E) A_{\delta}\left[\phi_{\delta} \chi_{\delta j}\right] d \tau, \nonumber \\
	K_{\gamma i, \mu} & =\int A_{\gamma}\left[\phi_{\gamma}^{\dagger} \chi_{\gamma i}\right](H-E) \Omega_{\mu, t} d \tau, \nonumber \\
	K_{\nu, \mu} & =\int \tilde{\Omega}_{\bar{\nu}, t}^{\dagger}(H-E) \Omega_{\mu, t} d \tau, \nonumber \\
	L_{\gamma i, \delta j} & = K_{\gamma i, \delta j}- K_{\gamma 0, \delta j}- K_{\gamma i, \delta 0}+ K_{\gamma 0, \delta 0}, \nonumber \\
	L_{\nu, \delta j} & = K_{\nu, \delta j}-K_{\nu, \delta 0}, \nonumber \\
	L_{\gamma i, \mu} & = K_{\gamma i, \mu}-K_{\gamma 0, \mu}, \nonumber \\
	L_{\nu, \mu} & = K_{\nu, \mu}, \nonumber \\
	M_{\gamma i}^{(c)} & =-K_{\gamma i, c 0}+K_{\gamma 0, c 0}, \quad(c=\alpha, \beta), \nonumber \\
	M_{\nu}^{(c)} & =-K_{\nu, c 0}, \quad(c=\alpha, \beta).
\end{align}

By solving the linear system of Eqs.~(\ref{a14}) and (\ref{a15}), we can obtain \(C_{\gamma i}^{(c)}\) and \(b_{\nu}^{(c)}\), and substitute them into Eq.~(\ref{a13}) to obtain the approximate $S$-matrix element \(S_{\gamma c, t}\). 
Finally, we obtain the stable $S$-matrix element \(S_{\beta \alpha, s t}\):
\begin{equation}
	S_{\beta \alpha, s t}=S_{\beta \alpha, t}-\frac{i k^{2}}{\hbar}\left[\sum_{\gamma} \sum_{i=0}^{n_{\gamma}} K_{\beta 0, \gamma i} C_{\gamma i}^{(\alpha)}+\sum_{\nu} K_{\beta 0, \nu} b_{\nu}^{(\alpha)}\right].
\end{equation}
Once we have obtained the $S$-matrix element, we can further calculate the scattering phase shift. 
Let's discuss two different cases. 
Take five-channel coupling as an example. 
If only one of the five channels is an open channel and the others are closed channels, then we study the scattering phase shift of the open channel after being affected by the closed channels. 
After using the above method to obtain the scattering matrix element of the open channel, we can substitute it into formula $S_L = |S_L|~ \text{exp}(2i\delta_L)$ to obtain the scattering phase shift. 
If we have more than one open channel in our study, we can use the above method to obtain twenty-five matrix elements, namely $S_{11}$, $S_{12}$, $S_{13}$, $\dots$, $S_{55}$ and then form a $5\times5$ matrix. 
Diagonalize the matrix to get five eigenvalues, which are the $S$ matrix elements of the five channels. 
Substitute them into formula $S_L = |S_L|~ \text{exp}(2i\delta_L)$ to get the scattering phase shifts of the five channels.

\end{document}